\documentstyle[prd,aps,preprint,tighten,epsfig]{revtex}

\begin{document}

\title{Flavor $S^{}_4 \otimes Z^{}_2$ symmetry and neutrino mixing}

\author{{\bf He Zhang} \thanks{E-mail: zhanghe@mail.ihep.ac.cn}}

\address{Institute of High Energy Physics, Chinese Academy of
Sciences,
\\ P.O. Box 918, Beijing 100049, China}

\maketitle

\begin{abstract}
We present a model of the lepton masses and flavor mixing based on
the discrete group $S^{}_4\otimes Z^{}_2$. In this model, all the
charged leptons and neutrinos are assigned to the ${\bf
\underline{3}}_\alpha$ representation of $S^{}_4$ in the Yamanouchi
bases. The charged lepton and neutrino masses are mainly determined
by the vacuum expectation value structures of the Higgs fields. A
nearly tri-bimaximal lepton flavor mixing pattern, which is in
agreement with the current experimental results, can be accommodated
in our model. The neutrino mass spectrum takes the nearly degenerate
pattern, and thus can be well tested in the future precise
experiments.
\end{abstract}

\newpage

\section{introduction}

To understand the origin of fermion masses and flavor mixing is
crucial and essential in modern particle physics. From the
analyses\cite{Vissani} of recent neutrino oscillation
experiments\cite{SNO,SK,KM,K2K,CHOOZ}, we have confirmed the large
solar mixing angle $\theta^{}_{12} \simeq 34^{\circ} $\cite{MSW},
maximal atmospheric mixing angle $\theta^{}_{23} \simeq
45^{\circ}$ and a very tiny $\theta^{}_{13}$, $\theta^{}_{13} <
10^{\circ}$. A very small solar mass squared difference $\Delta
m^2_{21} \equiv m^2_2-m^2_1 \simeq 8.0 \times 10^{-5}~ {\rm eV}^2$
and an atmospheric mass squared difference $\Delta m^2_{32} \equiv
m^2_3-m^2_2 \simeq \pm2.5 \times 10^{-3}~ {\rm eV}^2$ are given by
the experimental data\cite{Vissani}. Here the plus and minus signs
in front of $\Delta m^2_{32}$ correspond to the normal mass
hierarchy ($m^{}_1<m^{}_2<m^{}_3$) and inverted mass hierarchy
($m^{}_3<m^{}_1<m^{}_2$) case. Since the standard model predicts
massless neutrinos, it is clear that we should extend the standard
model to accommodate non-vanishing neutrino masses. Among all
possible mechanisms, the seesaw mechanism\cite{seesaw} is a very
interesting and elegant one to explain the light neutrino masses.
In the framework of the seesaw mechanism, the light left-handed
neutrino masses can well be understood by introducing heavy
right-handed Majorana neutrinos, and the light neutrino mass
matrix is given by
\begin{eqnarray}
M^{}_\nu=M^{}_D M^{-1}_R M^{T}_D \ ,
\end{eqnarray}
where $M^{}_D$ and $M^{}_R$ are the Dirac and Majorana mass matrices
respectively. The typical mass scales of $M^{}_R$ are $10^{14} \sim
10^{16} {\rm GeV}$.

Since the Yukawa structures are not well constrained up to now, to
identify the structure of the neutrino mass matrix is one of the
main objects in neutrino physics. An interesting and natural way
to study the Yukawa coupling and find the underlying physics is
the flavor symmetry. Among all possible flavor symmetries,
discrete non-Abelian groups such as $S^{}_3$\cite{S3},
$A^{}_4$\cite{A4} and so on have attracted a lot of attention.
Note that most of these models rely on the basis of their group
representation, and contain more theoretical parameters than the
observables.

The permutation group $S^{}_4$, which is formed by the $4!$
permutations, totally contains 24 group elements which belong to
five conjugate classes. Therefore it has five irreducible
representations ({\it reps}). Among these irreducible {\it reps},
there are two one-dimensional (${\bf \underline{1}}_S$ and ${\bf
\underline{1}}_A$), one two-dimensional (${\bf \underline{2}}$) and
two three-dimensional (${\bf \underline{3}}_\alpha$ and ${\bf
\underline{3}}_\beta$) {\it reps}. Here the `S' and `A' mean
symmetric and antisymmetric {\it reps} respectively. The character
table of $S^{}_4$ is given in Table 1. In respect that the particles
in our model are all assigned to the ${\bf \underline{1}}_S$, ${\bf
\underline{2}}$ and ${\bf \underline{3}}_\alpha$, we list the
relevant representation matrices in Appendix A.

The $S^{}_4$ models for fermion masses have been discussed by
several authors in Refs.\cite{S4,S4new}. In this note, we adopt the
$S^{}_4 \otimes Z^{}_2$ flavor symmetry and assign all the charged
leptons, neutrinos and Higgs into the irreducible Yamanouchi
bases\cite{chen}. Such a compact scheme contains only a few
parameters, and therefore it can be examined quite well in the
future experiments.

In the following, we will present the main contents of our model in
section II. Some detailed analytical and numerical analyses will be
given in section III. In section IV, the $S_4$ invariant Higgs
potential are discussed. Finally, a brief summary is given in
section V.

\section{particle assignment}

In our model, all the leptons are assigned to the $\underline{\bf
3}^{}_{\alpha}$ {\it rep} of $S^{}_4$, but their $Z^{}_2$ charges
are different. Under the group
\begin{eqnarray}
G=SU(3)^{}_c \otimes SU(2)^{}_L \otimes U(1)^{}_Y \otimes S^{}_4
\otimes Z^{}_2 \ ,
\end{eqnarray}
the lepton contents in our model are placed as
\begin{eqnarray}
\ell^{}_L & \sim & (1,2,-1)(\underline{\bf 3}^{}_{\alpha},+) \ ,
\nonumber
\\
e^{}_R & \sim & (1,1,-2)(\underline{\bf 3}^{}_{\alpha},+) \ ,
\nonumber
\\
\nu^{}_R & \sim & (1,1,0)(\underline{\bf 3}^{}_{\alpha},-) \ ,
\end{eqnarray}
where the plus or minus sign denotes the reflection properties
under $Z^{}_2$, i.e. $\Psi\rightarrow\pm\Psi$. The Higgs scalars
in our model are placed as follows
\begin{eqnarray}
H^{}_u & \sim & (1,2,-1)(\underline{\bf 1}^{}_{\bf S},+) \ ,
\nonumber \\
H^{}_d & \sim & (1,2,-1)(\underline{\bf 1}^{}_{\bf S},-) \ ,
\nonumber \\
\chi & \sim & (1,2,-1)(\underline{\bf 3}^{}_{\alpha},-) \ ,
\nonumber \\
\phi & \sim & (1,2,-1)(\underline{\bf 2}^{}_{},+) \ .
\end{eqnarray}
Here the $Z_2$ symmetry guarantees that the lepton doublet couples
to ($e_R, \phi_i$) and ($\nu_R, \chi_i$) respectively. It will be
shown in section IV that the masses of Higgs scalars are almost
unconstrained in this model, thus we can choose their masses at some
high energy scales in order to avoid the tree level flavor charging
neutral currents. Note that we do not introduce any $SU(2)^{}_L$
singlet or triplet Higgs\cite{triplet}. Hence the Heavy right-handed
neutrino masses are exactly degenerate.

By using the group algebra given in Appendix A, we write down the
$S^{}_4 \otimes Z^{}_2$ invariant Yukawa couplings:
\begin{eqnarray}
-{\cal L}^{}_{Y}=\alpha^{}_e \overline{\ell^{}_{iL}}
e^{}_{iR}\tilde{H^{}_u}+\beta^{}_e f^{}_{ijk}
\overline{\ell^{}_{iL}} e^{}_{jR}\tilde{\phi^{}_k} +
\alpha^{}_{\nu} \overline{\ell^{}_{iL}} \nu^{}_{iR}{H^{}_d}
+\beta^{}_\nu g^{}_{ijk} \overline{\ell^{}_{iL}}
\nu^{}_{jR}{\chi^{}_k} + \frac{1}{2}M\overline{\nu^{c}_{iR}}
\nu^{}_{iR}+h.c.
\end{eqnarray}
where $\tilde{H}^{}_u=i\tau^{}_2H^{*}_u$ and
$\tilde{\phi}=i\tau^{}_2\phi^{*}$. The last term in Eq. (5) is the
bare Majorana mass term with $M$ being the typical mass scale of
the heavy right-handed neutrinos. The coefficients
$\alpha^{}_{e,\nu}$ and $\beta^{}_{e,\nu}$ are in general complex
parameters. However, since their phases are all global phases,
there will be no Dirac CP violating phases in the MNS matrix and
only the Majorana phases can be accommodated up to now.
\footnote{Here we do not consider the spontaneous CP
violation\cite{SCP}}
The structures of the traceless matrices $f$ and $g$ can be
obtained from the CG coefficients of $S^{}_4$. Following the CG
coefficient tables given in Ref. \cite{chen}, we arrive at
\begin{eqnarray}
f^{}_{ij1}=\left(\matrix{ 0 & \sqrt{2} & 0 \cr \sqrt{2} & 1 & 0
\cr 0 & 0 & -1 }\right) \ , \ \ \ \ \ f^{}_{ij2}=\left(\matrix{ 0
& 0 & \sqrt{2} \cr 0 & 0 & -1 \cr \sqrt{2} & -1 & 0}\right) \ ,
\end{eqnarray}
and
\begin{eqnarray}
g^{}_{ij1}&=&\left(\matrix{ 2 & 0 & 0 \cr 0 & -1 & 0 \cr 0 & 0 &
-1 }\right) \ , \ \ \ \ \ g^{}_{ij2}=\left(\matrix{ 0 & -1 & 0 \cr
-1 & \sqrt{2} & 0 \cr 0 & 0 & -\sqrt{2}}\right) \ , \nonumber \\
g^{}_{ij3}&=&\left(\matrix{ 0 & 0 & -1 \cr 0 & 0 & -\sqrt{2} \cr
-1 & -\sqrt{2} & 0 }\right) \ ,
\end{eqnarray}
where all the coefficient matrices are symmetric and traceless.
Hence both the charged lepton and neutrino mass matrices are
symmetric.

As we have mentioned above, the right-handed neutrino masses are
exactly degenerate by now. Such a result is not favored in some
leptogenesis models\cite{leptogenesis} in which the mass splits of
the heavy right-handed neutrinos are required. This drawback can
easily be solved by introducing an additional Higgs scalar $\eta
\sim (1,1,0)(\underline{\bf 2}^{}_{},+)$ or $\xi \sim
(1,1,0)(\underline{\bf 3}^{}_{\alpha},+)$ which couples to
$\nu^{}_R$ in the form of
\begin{eqnarray}
-{\cal L}^{}_{\eta-\nu^{}_R} = \beta^{}_R f^{}_{ijk}
\overline{\nu^{c}_{iR}} \nu^{}_{jR} \eta^{}_k \ ,
\end{eqnarray}
or
\begin{eqnarray}
-{\cal L}^{}_{\xi-\nu^{}_R} = \beta^{}_R g^{}_{ijk}
\overline{\nu^{c}_{iR}} \nu^{}_{jR} \xi^{}_k \ .
\end{eqnarray}
Their vacuum expectation values(VEVs) will bring the mass splits
of right-handed neutrinos and the phases of $\beta^{}_R$ will lead
to the Dirac CP violating phases in the MNS matrix.

Another possible way to gain the mass differences of the
right-handed neutrinos is to consider the renormalization group
equation running effects\cite{RGE} which can usually be used to
generate a small mass split in the resonant leptogenesis
models\cite{resonant}. A Dirac CP violating phase can also be
gained at the same time from the renormalization group equation
running\cite{Dirac}.

In our analyses, we focus our attention on the low energy
phenomena of our model. So we assume that all the Yukawa couplings
are real and the right-handed neutrino masses are exactly
degenerate.

\section{lepton mass matrices and mixing matrix}

Assuming the flavor symmetry $S^{}_4$ is broken by the VEVs of the
Higgs scalars in the form of $\left<H^{}_u\right>=v^{}_u$,
$\left<H^{}_d\right>=v^{}_d$, $\left<\phi^{}_i\right>=u^{}_i$ and
$\left<\chi^{}_i\right>=w^{}_i$, we obtain the mass matrix of the
charged leptons as
\begin{eqnarray}
M^{}_l=\left( \matrix{a & \sqrt{2} x^{}_1 & \sqrt{2} x^{}_2 \cr
\sqrt{2}x^{}_1 & a+ x^{}_1 & -x^{}_2 \cr \sqrt{2} x^{}_2 & -x^{}_2
& a - x^{}_1 } \right) \ ,
\end{eqnarray}
where $a$ and $x^{}_i$ are defined by $a=\alpha^{}_{e}v^{}_u$ and
$x^{}_i=\beta^{}_{e} u^{}_i$ respectively.

The Dirac mass matrix of the neutrinos is given by
\begin{eqnarray}
M^{}_D=\left(\matrix{ 2 y^{}_1+b & -y^{}_2 & -y^{}_3 \cr -y^{}_2 &
\sqrt{2} y^{}_2 -y^{}_1+b & -\sqrt{2}y^{}_{3} \cr -y^{}_3 &
-\sqrt{2}y^{}_{3} & -y^{}_1-\sqrt{2}y^{}_2+b}\right) \ ,
\end{eqnarray}
where $b=\alpha^{}_\nu v^{}_d$ and $y^{}_i=\beta^{}_\nu w^{}_i$.
Taking account of the seesaw relation in Eq. (1), we obtain the
light neutrino mass matrix:
\begin{eqnarray}
M^{}_\nu=m^{}_0\left(\matrix{ (2+\epsilon)^2+r^2+s^2 &
\sqrt{2}s^2-(1+2\epsilon+\sqrt{2}r)r & -(1+2\epsilon-2\sqrt{2}r)s
\cr \sim & (1 - \epsilon - \sqrt{2}r)^2+r^2+ 2 s^2 & (2\sqrt{2}
-2\sqrt{2}\epsilon + r)s \cr \sim & \sim &
(1-\epsilon+\sqrt{2}r)^2+3s^2}\right) \ ,
\end{eqnarray}
where $m^{}_0=y^2_1/M$, $\epsilon=b/y^{}_1$, $r=y^{}_2/y^{}_1$ and
$s=y^{}_3/y^{}_1$. Note that all the lepton mass matrices are
symmetric.

The mass matrices $M^{}_l$ and $M^{}_\nu$ can respectively be
diagonalized by two unitary matrices $V^{}_l$ and $V^{}_\nu$:
$V^{\dagger}_l M^{}_l V^{\ast}_l={\rm Diag}\{ m^{}_e,
m^{}_\mu,m^{}_\tau \}$, $V^{\dagger}_\nu M^{}_\nu
V^{\ast}_\nu={\rm Diag}\{ m^{}_1, m^{}_2,m^{}_3 \}$. The lepton
flavor mixing matrix $V^{}_{\rm MNS}$, which links the neutrino
mass eigenstates ($\nu^{}_1,\nu^{}_2,\nu^{}_3$) with their flavor
eigenstates ($\nu^{}_e,\nu^{}_\mu,\nu^{}_\tau$), is then given by
$V^{}_{\rm MNS}=V^{\dag}_{l}V^{}_{\nu}$.

\subsection{charged lepton sector}

By diagonalizing the charged lepton mass matrix in Eq. (10), we get
the charged lepton masses:
\begin{eqnarray}
m^{}_e & = & a+2x^{}_1 \ , \nonumber \\
m^{}_\mu & = & a-x^{}_1+\sqrt{3}x^{}_2 \ , \nonumber \\
m^{}_\tau & = & a-x^{}_1-\sqrt{3}x^{}_2  \ ,
\end{eqnarray}
and the mixing matrix $V^{}_l$:
\begin{eqnarray}
V^{}_l = \left(\matrix{\displaystyle \frac{1}{\sqrt{3}} &
\displaystyle \frac{1}{\sqrt{3}} & \displaystyle
\frac{1}{\sqrt{3}}  \cr \displaystyle \frac{2}{\sqrt{6}} &
\displaystyle -\frac{1}{\sqrt{6}} & \displaystyle
-\frac{1}{\sqrt{6}}  \cr 0 & \displaystyle \frac{1}{\sqrt{2}}
&\displaystyle -\frac{1}{\sqrt{2}}  }\right) \ .
\end{eqnarray}
A typical interesting and instructive feature is that $V^{}_l$ is a
constant matrix with respective to arbitrary $a$ and $x^{}_{i}$.
After being transposed, this matrix takes the very similar form to
the tri-bimaximal mixing matrix\cite{tribi}
\begin{equation}
V^{}_{\rm tri-bi} = \left ( \matrix{ \displaystyle
\frac{2}{\sqrt{6}} & \displaystyle \frac{1}{\sqrt{3}} & 0 \cr\cr
\displaystyle -\frac{1}{\sqrt{6}} & \displaystyle
\frac{1}{\sqrt{3}} & \displaystyle \frac{1}{\sqrt{2}} \cr\cr
\displaystyle -\frac{1}{\sqrt{6}} & \displaystyle
\frac{1}{\sqrt{3}} & \displaystyle -\frac{1}{\sqrt{2}} \cr} \right
)  \; ,
\end{equation}
only up to an exchange between the first and second column.

By using Eq. (13), we can also express the parameters $a$, $x^{}_1$
and $x^{}_2$ in terms of the charged lepton masses as
\begin{eqnarray}
a & = & \frac{1}{3}\left( m^{}_e+m^{}_\mu+m^{}_\tau \right) \ ,
\nonumber \\
x^{}_1 & = & \frac{1}{6}\left( 2m^{}_e-m^{}_\mu-m^{}_\tau \right) \ ,\nonumber \\
x^{}_2 & = & \frac{1}{2\sqrt{3}}\left( m^{}_\mu-m^{}_\tau \right)
\ .
\end{eqnarray}
Considering the mass hierarchies of the charged leptons:
$m^{}_\tau\gg m^{}_\mu \gg m^{}_e$, it is easy to conclude that
$|a|>|x^{}_2|>|x^{}_1|$.

\subsection{neutrino sector}

As an example, we assume $s=0$ and $r \ll 1$
\footnote{In the case of $s \neq 0$, there will be more freedom for
us to fit the data in principle. However, it will involve the
problem of cubic root. Since the main propose of this paper is to
show the feasibility of this $S^{}_4$ model, we simply take $s=0$ as
an example. In our numerical calculations at the end of section III,
we do not make any approximation on $r$ or $s$, and we will see that
the exact numerical result satisfies the condition $s,r \ll 1$ very
well.}.
Then, $M^{}_D$ takes the simple form:
\begin{eqnarray}
M^{}_D=y^{}_1 \left(\matrix{ 2 +\epsilon & -r & 0 \cr -r &
-1+\sqrt{2} r +\epsilon & 0 \cr 0 & 0 &
-1-\sqrt{2}r+\epsilon}\right) \ ,
\end{eqnarray}
and the light neutrino mass matrix can be written as
\begin{eqnarray}
M^{}_\nu = m^{}_0 \left(\matrix{ (2+\epsilon)^2+r^2 &
-r-2r\epsilon-\sqrt{2}r^2 & 0 \cr -r-2r\epsilon-\sqrt{2}r^2 &
(1-\epsilon-\sqrt{2}r)^2+r^2 & 0 \cr 0 & 0 &
(1-\epsilon+\sqrt{2}r)^2 }\right) \ .
\end{eqnarray}
There are two texture zeros in the third row of
$M^{}_\nu$\cite{zero}. Therefore only the mixing between the first
two generations exists in the neutrino sector. After straightforward
calculations, we obtain the light neutrino masses:
\begin{eqnarray}
m^{}_1 & \simeq &
m^{}_0\left[(1-\epsilon)^2-2\sqrt{2}(1-\epsilon)r + \frac{2}{3}
(4-\epsilon)r^2\right]  \ , \nonumber \\
m^{}_2 & \simeq & m^{}_0 \left[ (2 + \epsilon)^2 + \frac{2}{3} ( 2 +
\epsilon) r^2 \right]  \ , \nonumber \\ m^{}_3 & \simeq &
m^{}_0\left[ (1-\epsilon)^2+2\sqrt{2}(1-\epsilon)r+2r^2 \right] \ ,
\end{eqnarray}
where the terms in proportion to $r^3$ have been neglected. The
mixing matrix $V^{}_\nu$ can be approximated to
\begin{eqnarray}
V^{}_\nu \simeq \left(\matrix{ \displaystyle
\frac{1}{3}r+\frac{\sqrt{2}}{9}r^2 & \displaystyle
-1+\frac{1}{18}r^2 & 0 \cr \displaystyle 1-\frac{1}{18}r^2 &
\displaystyle \frac{1}{3}r+\frac{\sqrt{2}}{9}r^2 & \displaystyle 0
\cr 0 & 0 & 1 }\right) + {\cal O}(r^3) \ .
\end{eqnarray}
It is interesting that only the parameter $r$ appears in the
approximate expression of $V^{}_\nu$.

The MNS mixing matrix is given by
\begin{eqnarray}
V^{}_{\rm MNS} & = &  V^\dagger_l V^{}_\nu = \left [ V^\dagger_l .
\left(\matrix{0 & 1 & 0 \cr 1 & 0 & 0 \cr 0 & 0 & 1}\right)
\right] \cdot \left[ \left(\matrix{0 & 1 & 0 \cr 1 & 0 & 0 \cr 0 &
0 & 1}\right) V ^{}_\nu \right ] \nonumber \\
& = &  \left ( \matrix{ \displaystyle \frac{2}{\sqrt{6}} &
\displaystyle \frac{1}{\sqrt{3}} & 0 \cr\cr \displaystyle
-\frac{1}{\sqrt{6}} & \displaystyle \frac{1}{\sqrt{3}} &
\displaystyle \frac{1}{\sqrt{2}} \cr\cr \displaystyle
-\frac{1}{\sqrt{6}} & \displaystyle \frac{1}{\sqrt{3}} &
\displaystyle -\frac{1}{\sqrt{2}} \cr} \right ) \cdot
\left(\matrix{  \displaystyle 1-\frac{1}{18}r^2 & \displaystyle
\frac{1}{3}r+\frac{\sqrt{2}}{9}r^2 & \displaystyle 0 \cr
\displaystyle \frac{1}{3}r+\frac{\sqrt{2}}{9}r^2 &
\displaystyle -1+\frac{1}{18}r^2 & 0 \cr 0 & 0 & 1 }\right)\nonumber \\
&=& V^{}_{\rm tri-bi} \cdot \left(\matrix{ \cos\theta & \sin\theta
& 0 \cr -\sin\theta & \cos\theta & 0 \cr 0 & 0 & 1 }\right)\cdot
\left(\matrix{1 & 0 & 0 \cr 0 & -1 & 0 \cr 0 & 0 & 1}\right) \ ,
\end{eqnarray}
where
\begin{eqnarray}
\sin\theta & = &-\frac{1}{3}r-\frac{\sqrt{2}}{9}r^2  \ , \\
\cos\theta & = &1-\frac{1}{18}r^2 \ .
\end{eqnarray}
The last matrix in Eq. (21) contributes a trivial phase $\pm\pi$
in the MNS matrix, thus it can be safely neglected without
changing any phisics. The explicit expression of $V^{}_{\rm MNS}$
is
\begin{eqnarray}
V^{}_{\rm MNS}=\left(\matrix{\displaystyle
\frac{2\cos\theta}{\sqrt{6}}-\frac{\sin\theta}{\sqrt{3}} &
\displaystyle
\frac{\cos\theta}{\sqrt{3}}+\frac{2\sin\theta}{\sqrt{6}} &
 0 \cr
\displaystyle \frac{-\cos\theta}{\sqrt{6}} -
\frac{\sin\theta}{\sqrt{3}} & \displaystyle
\frac{\cos\theta}{\sqrt{3}}-\frac{\sin\theta}{\sqrt{6}} &
\displaystyle \frac{1}{\sqrt{2}} \cr \displaystyle
-\frac{\cos\theta}{\sqrt{6}} - \frac{\sin\theta}{\sqrt{3}}
&\displaystyle \frac{\cos\theta}{\sqrt{3}} -
\frac{\sin\theta}{\sqrt{6}} & \displaystyle
-\frac{1}{\sqrt{2}}}\right) \ .
\end{eqnarray}
Comparing Eq. (24) with the standard parametrization
\begin{eqnarray}
V^{\rm sp}_{\rm MNS}  =  \left( \matrix{c^{}_{12} c^{}_{13} &
s^{}_{12} c^{}_{13} & s^{}_{13} e^{-i\delta} \cr -s^{}_{12}
c^{}_{23}-c^{}_{12} s^{}_{23} s^{}_{13}
 e^{i \delta} & c^{}_{12} c^{}_{23}-s^{}_{12} s^{}_{23} s^{}_{13}
 e^{i \delta} & s^{}_{23} c^{}_{13} \cr
 s^{}_{12} c^{}_{23}-c^{}_{12} c^{}_{23} s^{}_{13}
 e^{i \delta} & -c^{}_{12} s^{}_{23}-s^{}_{12} c^{}_{23} s^{}_{13}
 e^{i \delta} & c^{}_{23} c^{}_{13}}
\right) \cdot \left(\matrix{e^{i\rho} & 0 & 0 \cr 0 & e^{i \sigma}
& 0 \cr 0 & 0 & 1}\right) \ ,
\end{eqnarray}
we obtain three mixing angles
\begin{eqnarray}
\sin\theta^{}_{12} & = & \frac{\cos\theta}{\sqrt{3}} +
\frac{2\sin\theta}{\sqrt{6}} \ ,
\nonumber \\
\sin\theta^{}_{23} & = & \frac{1}{\sqrt{2}} \ , \nonumber \\
\sin\theta^{}_{13} & = & 0 \ ,
\end{eqnarray}
together with $\delta=\rho=\sigma=0$.

From the experimental data of $\theta^{}_{12}$ and Eq. (26), we can
estimate that $\theta \simeq -1.4^{\circ}$. These results depend on
ours assumptions. It should be emphasized that $\theta^{}_{13}$ will
not vanish if we loose our assumptions by setting $s \neq 0$. In
this case, small corrections on $\theta^{}_{23}$ will also be given
corresponding to the nonzero $s$, and we will illustrate these
corrections in the next subsection.

\subsection{numerical analyses}

In our numerical analyses, we first take the charged lepton pole
masses $M^{\rm pole}_l$ given by PDG\cite{PDG}, and then evaluate
the charged lepton running masses $m^{}_l(\mu)$ from $M^{\rm
pole}_l$ to the $M^{}_Z$ scale by using\cite{running}
\begin{eqnarray}
m^{}_l(M^{}_Z) \; = \; M^{\rm pole}_l \left [1 -
\frac{\alpha(M^{}_Z)}{\pi} \left ( \frac{3}{2} \ln
\frac{M^{}_Z}{m^{}_l(M^{}_Z)} + 1 \right )\right ] \ .
\end{eqnarray}
Then we obtain
\begin{eqnarray}
m^{}_e(M^{}_Z) & = & 0.4867 ~{\rm MeV} \ , \nonumber \\
m^{}_\mu(M^{}_Z) & = & 102.7 ~{\rm MeV} \ , \nonumber \\
m^{}_\tau(M^{}_Z) & = & 1746.6 ~{\rm MeV} \ .
\end{eqnarray}
We can directly calculate the parameters ($a, x_1, x_2$) by using
Eqs. (16),
\begin{eqnarray}
a & \simeq & 616.6 ~ {\rm MeV} \ , \ \ \ \ \ x^{}_1  \simeq -308.0 ~
{\rm MeV} \ , \ \ \ \ \ x^{}_2 \simeq -474.5 ~ {\rm MeV} \ .
\end{eqnarray}

For the neutrino sector, we take the light neutrino mass squared
differences $\Delta m^2_{21}
 = 8.0 \times 10^{-5}~ {\rm eV}^2$ and $\Delta m^2_{32} = \pm 2.5 \times
10^{-3} ~{\rm eV}^2$, three mixing angles
$\theta^{}_{12}=33.9^{\circ}$, $\theta^{}_{23}=45^{\circ}$ and
$\theta^{}_{13}=0^{\circ}$\cite{Vissani} as the input values. We
firstly consider the case $s=0$. By combining Eqs. (19), (22) and
(23), after some numerical calculations, we find that only the
normal mass hierarchy case is allowed for our assumptions. The
corresponding parameters are
\begin{eqnarray}
m^{}_0 & \simeq & 2.99\times 10^{-2} ~ {\rm eV} \ , \ \ \ \ \ r
\simeq 6.92 \times 10^{-2} \ , \ \ \ \ \ \epsilon \simeq -0.545 \ .
\end{eqnarray}
It can be seen clearly that our assumption $r \ll 1$ is quite
reasonable. The lightest neutrino mass is $m^{}_1 \simeq 0.063 ~{\rm
eV}$. Hence the neutrinos have a nearly degenerate mass spectrum:
$m^{}_1 \simeq m^{}_2 \simeq m^{}_3$. As for the case $s\neq 0$, the
mixing angles $\theta_{13}$ and $\theta_{23}$ will in general get
corrections and deviate from their tribi-maximal values. In Fig. 1,
we show the corrections on $\theta_{13}$ and $\theta_{23}$ from
nonzero $s$. We can find that $\theta_{13}$ is more sensitive to the
non-vanishing $s$, and in order to get a nonzero $\theta_{13}$
within 1 and 2$\sigma$ confidence level $s$ should not be larger
than 0.02 and 0.035.

Note that the predictions of our model rely on the VEV structures of
the Higgs fields. Here we have assumed that such conditions on the
VEVs of the Higgs scalars can be well satisfied. In section IV, a
detailed analysis will be given to show that this kind of VEV
configuration can be obtained by a suitable choice of parameters of
the Higgs potential. As mentioned above, in our analyses we do not
consider the CP violation of our model. To generalize our model to
the CP violating case, we can add the singlet Higgs scalar given by
Eq. (8) or (9) and insert the phase factors into the parameters
$\alpha$ and $\beta$ in Eq. (5). In this case, the mass splits of
the right-handed neutrinos and the CP violation in the neutrino
mixing can be acquired simultaneously. As a consequence, there will
be more freedom to fit the experimental data and one may test the
leptogenesis mechanism within this framework. Such an extension may
be more interesting and the corresponding detailed analyses will be
elaborated elsewhere.

\section{higgs potential}

The most general $ S_4 \otimes Z_2 $ invariant Higgs potential in
our model is given by
\begin{eqnarray}
V_{} & = & -m^2_u (H^{\dagger}_u H_u) -m^2_d (H^{\dagger}_d H_d)
-m^2_\phi \sum^2_{i=1}\phi^{\dagger}_i \phi_i -m^2_\chi
\sum^3_{j=1}\chi^{\dagger}_j \chi_j \nonumber \\
& + & \lambda_u(H^{\dagger}_u H_u)^2 +\lambda_d(H^{\dagger}_d H_d)^2
+ \lambda_\phi \left(\sum^2_{i=1}\phi^{\dagger}_i \phi_i\right)^2
+\lambda_\chi
\left( \sum^3_{j=1}\chi^{\dagger}_j \chi_j\right)^2 \nonumber \\
& + &  \tau_1 \left[ (\phi^{\dagger}_1 \phi_1 - \phi^{\dagger}_2
\phi_2)^2 + (\phi^{\dagger}_1 \phi_2 + \phi^{\dagger}_2 \phi_1)^2
\right ] + \tau_2 (\phi^{\dagger}_1 \phi_2 - \phi^{\dagger}_2
\phi_1)^2 \nonumber \\
& + & \mu_1 \left[ (2\chi^\dagger_1 \chi_1 -\chi^\dagger_2 \chi_2
-\chi^\dagger_3 \chi_3)^2 + (\chi^\dagger_1 \chi_2 + \chi^\dagger_2
\chi_1 - \sqrt{2} \chi^\dagger_2 \chi_2 +
\sqrt{2} \chi^\dagger_3 \chi_3)^2 \right. \nonumber \\
& + & \left. (\chi^\dagger_1 \chi_3 + \chi^\dagger_3 \chi_1 +
\sqrt{2} \chi^\dagger_2 \chi_3 + \sqrt{2}
\chi^\dagger_3 \chi_2 )^2\right] \nonumber \\
& + & \mu_2 \left[ (\chi^\dagger_1 \chi_2 - \chi^\dagger_2 \chi_1)^2
+ (\chi^\dagger_1 \chi_3 - \chi^\dagger_3 \chi_1)^2  +
(\chi^\dagger_2 \chi_3 - \chi^\dagger_3 \chi_2)^2 \right] \nonumber
\\
& + & \mu_3  \left[ (\chi^\dagger_2 \chi_2 - \chi^\dagger_3 \chi_3 +
\sqrt{2} \chi^\dagger_1 \chi_2 + \sqrt{2} \chi^\dagger_2 \chi_1 )^2
+  (\chi^\dagger_2 \chi_3 + \chi^\dagger_3 \chi_2 - \sqrt{2}
\chi^\dagger_1 \chi_3 - \sqrt{2} \chi^\dagger_3 \chi_1 )^2 \right] \nonumber \\
& + & \lambda_1 (H^{\dagger}_u H_u)(H^{\dagger}_d H_d) + \lambda_2
(H^{\dagger}_u H_d)(H^{\dagger}_d H_u) + \left [ \lambda_3
(H^{\dagger}_u H_d)^2+{\rm h.c.} \right] \nonumber \\
& + & \lambda^\phi_1 (H^{\dagger}_u
H_u)\left(\sum^2_{i=1}\phi^{\dagger}_i \phi_i\right) +
 \left\{ \lambda^\phi_2 \left[ ( H^{\dagger}_u \phi_1)^2
+ (H^{\dagger}_u \phi_2)^2\right] + {\rm h.c.}\right\}
+\lambda^\phi_3 \left( \left|H^{\dagger}_u \phi_1 \right|^2 +
\left|H^{\dagger}_u
\phi_2 \right|^2 \right) \nonumber \\
& + & \left\{\lambda^\phi_4 \left[ (H^{\dagger}_u \phi_1
)(\phi^{\dagger}_1 \phi_1 - \phi^{\dagger}_2 \phi_2) -
(H^{\dagger}_u \phi_2 )(\phi^{\dagger}_1 \phi_2 + \phi^{\dagger}_2
\phi_1) \right] +{\rm h.c. }\right\} \nonumber \\
& + & \lambda^\chi_1 (H^{\dagger}_u
H_u)\left(\sum^3_{i=1}\chi^{\dagger}_i \chi_i\right) + \left\{
\lambda^\chi_2 \left[ ( H^{\dagger}_u \chi_1)^2 + (H^{\dagger}_u
\chi_2)^2 + (H^{\dagger}_u \chi_3)^2\right] + {\rm h.c.}\right\}
\nonumber \\
& + & \lambda^\chi_3 \left( \left|H^{\dagger}_u \chi_1 \right|^2 +
\left|H^{\dagger}_u \chi_2 \right|^2 +\left|H^{\dagger}_u \chi_3
\right|^2 \right) \nonumber \\
& + & \sigma^\phi_1 (H^{\dagger}_d
H_d)\left(\sum^2_{i=1}\phi^{\dagger}_i \phi_i\right) +
 \left\{ \sigma^\phi_2 \left[ ( H^{\dagger}_d \phi_1)^2
+ (H^{\dagger}_d \phi_2)^2\right] + {\rm h.c.}\right\}
+\sigma^\phi_3 \left( \left|H^{\dagger}_d \phi_1 \right|^2 +
\left|H^{\dagger}_d
\phi_2 \right|^2 \right) \nonumber \\
& + & \sigma^\chi_1 (H^{\dagger}_d
H_d)\left(\sum^3_{i=1}\chi^{\dagger}_i \chi_i\right) + \left\{
\sigma^\chi_2 \left[ ( H^{\dagger}_d \chi_1)^2 + (H^{\dagger}_d
\chi_2)^2 + (H^{\dagger}_d \chi_3)^2\right] + {\rm h.c.}\right\}
\nonumber \\
& + & \sigma^\chi_3 \left( \left|H^{\dagger}_d \chi_1 \right|^2 +
\left|H^{\dagger}_d \chi_2 \right|^2 +\left|H^{\dagger}_d \chi_3
\right|^2 \right) \nonumber \\
& + & \left\{ \sigma^\chi_4 \left[
(H^{\dagger}_d\chi_1)(2\chi^\dagger_1 \chi_1 -\chi^\dagger_2 \chi_2
-\chi^\dagger_3 \chi_3) - (H^{\dagger}_d\chi_2)(\chi^\dagger_1
\chi_2 + \chi^\dagger_2 \chi_1 - \sqrt{2} \chi^\dagger_2 \chi_2 +
\sqrt{2} \chi^\dagger_3 \chi_3) \right. \right. \nonumber \\
& - & \left.\left. (H^{\dagger}_d\chi_3)(\chi^\dagger_1 \chi_3 +
\chi^\dagger_3 \chi_1 + \sqrt{2} \chi^\dagger_2 \chi_3 + \sqrt{2}
\chi^\dagger_3 \chi_2 ) \right] + { \rm h.c. } \right\} \nonumber \\
& + & \kappa_1 \left(\sum^2_{i=1}\phi^{\dagger}_i
\phi_i\right)\left( \sum^3_{j=1}\chi^{\dagger}_j \chi_j\right) +
\left\{ \kappa_2 \left [ (\phi^{\dagger}_1 \phi_1 - \phi^{\dagger}_2
\phi_2)(\chi^\dagger_2 \chi_2 - \chi^\dagger_3 \chi_3 + \sqrt{2}
\chi^\dagger_1 \chi_2 + \sqrt{2} \chi^\dagger_2 \chi_1 )
\right.\right. \nonumber \\
& + & \left.\left. (\phi^{\dagger}_1 \phi_2 + \phi^{\dagger}_2
\phi_1)(\chi^\dagger_2 \chi_3 + \chi^\dagger_3 \chi_2 - \sqrt{2}
\chi^\dagger_1 \chi_3 - \sqrt{2} \chi^\dagger_3 \chi_1 ) \right]
+{\rm h.c.} \right\} \nonumber \\
& + & \left\{ \kappa_3 \left[ 2(\phi^\dagger_1 \chi_2
+\phi^\dagger_2 \chi_3 )^2 + (\phi^\dagger_1 \chi_2 - \phi^\dagger_2
\chi_3 + \sqrt{2}\phi^\dagger_1 \chi_1)^2 + (\phi^\dagger_2 \chi_2 +
\phi^\dagger_1 \chi_3 -
\sqrt{2}\phi^\dagger_2 \chi_1)^2   \right] + {\rm h.c.}\right\} \nonumber \\
& + & \left\{ \kappa_4 \left[ (\phi^\dagger_1 \chi_2 -
\phi^\dagger_2 \chi_3 - \sqrt{2}\phi^\dagger_1 \chi_1)^2 +
(\phi^\dagger_2 \chi_2 + \phi^\dagger_1 \chi_3 +
\sqrt{2}\phi^\dagger_2 \chi_1)^2 + 2(\phi^\dagger_2 \chi_2 -
\phi^\dagger_1 \chi_3 )^2 \right] + {\rm h.c.}\right\} \nonumber \\
& + &  \kappa_5 \left[ 2\left|\phi^\dagger_1 \chi_2 +\phi^\dagger_2
\chi_3 \right|^2 + \left|\phi^\dagger_1 \chi_2 - \phi^\dagger_2
\chi_3 + \sqrt{2}\phi^\dagger_1 \chi_1\right|^2 +
\left|\phi^\dagger_2 \chi_2 + \phi^\dagger_1 \chi_3 -
\sqrt{2}\phi^\dagger_2 \chi_1\right|^2   \right]  \nonumber \\
& + &  \kappa_6 \left[ \left|\phi^\dagger_1 \chi_2 - \phi^\dagger_2
\chi_3 - \sqrt{2}\phi^\dagger_1 \chi_1\right|^2 +
\left|\phi^\dagger_2 \chi_2 + \phi^\dagger_1 \chi_3 +
\sqrt{2}\phi^\dagger_2 \chi_1\right|^2 + 2\left|\phi^\dagger_2
\chi_2 -
\phi^\dagger_1 \chi_3 \right|^2 \right] \nonumber \\
& + & \left\{ \delta_1 \left[ (H^{\dagger}_u \phi_1 )(\chi^\dagger_2
\chi_2 - \chi^\dagger_3 \chi_3 + \sqrt{2} \chi^\dagger_1 \chi_2 +
\sqrt{2} \chi^\dagger_2 \chi_1) \right.\right. \nonumber \\
& - & \left.\left. (H^{\dagger}_u \phi_2 )(\chi^\dagger_2 \chi_3 +
\chi^\dagger_3 \chi_2 - \sqrt{2} \chi^\dagger_1 \chi_3 - \sqrt{2}
\chi^\dagger_3 \chi_1) \right] +{\rm h.c.} \right\} \nonumber \\
& + & \left\{ \delta_2 \left[ \sqrt{2}(H^\dagger_u
\chi_1)(\phi^\dagger_1 \chi_2 +\phi^\dagger_2 \chi_3 ) +
(H^\dagger_u \chi_2) (\phi^\dagger_1 \chi_2 - \phi^\dagger_2 \chi_3
+ \sqrt{2}\phi^\dagger_1 \chi_1) \right.\right. \nonumber \\
& - & \left.\left. (H^\dagger_u \chi_3)(\phi^\dagger_1 \chi_3 +
\phi^\dagger_2 \chi_2 -
\sqrt{2}\phi^\dagger_2 \chi_1)   \right] + {\rm h.c.}\right\} \nonumber \\
& + & \left\{ \delta_3 \left[ \sqrt{2}(H^\dagger_u
\chi_1)(\chi^\dagger_2 \phi_1 +\chi^\dagger_3 \phi_2 ) +
(H^\dagger_u \chi_2) (\chi^\dagger_2 \phi_1 - \chi^\dagger_3 \phi_2
+ \sqrt{2}\chi^\dagger_1 \phi_1) \right.\right. \nonumber \\
& - & \left.\left. (H^\dagger_u \chi_3)(\chi^\dagger_3 \phi_1 +
\chi^\dagger_2 \phi_2 - \sqrt{2}\chi^\dagger_1 \phi_2)   \right] +
{\rm h.c.}\right\} \ .
\end{eqnarray}
The Higgs potential contains totally 39 parameters. Among these
parameters 12 of them are in general complex and the rest are real.
Here we take all the parameters and VEVs to be real and the minimum
of Higgs potential can be written as
\begin{eqnarray}
V_{\rm min} & = & -m^2_u v^2_u - m^2_d v^2_d - m^2_\phi(u^2_1 +
u^2_2) -  m^2_\chi(w^2_1 + w^2_2 + w^2_3) \nonumber \\
& + & \lambda_u v^4_u + \lambda_d v^4_d + (\lambda_\phi+\tau_1)(
u^4_1+ u^4_2 ) + (\lambda_\chi + 4 \mu_1)w^4_1 + (\lambda_\chi + 3
\mu_1 + \mu_3)(w^4_2 + w^4_3) \nonumber \\
& + & \ldots \ .
\end{eqnarray}
The full analytical formulae of $V_{\rm min}$ have been listed in
Table 2. Note that the couplings $\tau_2$ and $\mu_2$ do not appear
in the Higgs minimal. The number of parameters in the Higgs
potential is more than 30, thus we are confident that it is adequate
to arrive at the suitable minimum of the Higgs potential, and the
VEV structure in section III can be satisfied.

\section{summary}

We have presented a lepton mass model based on the discrete group
$S^{}_4\otimes Z^{}_2$. The right-handed charged leptons and
$SU(2)^{}_L$ doublets are all assigned to the ${\bf
\underline{3}}_\alpha$ {\it reps} of $S^{}_4$ with plus $Z^{}_2$
charges, and the heavy right-handed neutrinos are embedded in the
${\bf \underline{3}}_\alpha$ with minus $Z^{}_2$ charges. Four
sets of Higgs doublets are introduced and the lepton masses are
mainly determined by the VEV structures of the Higgs fields. After
diagonalizing the lepton mass matrices, we obtained a nearly
tri-bimaximal mixing matrix together with a nearly degenerate
light neutrino mass spectrum. By some careful analytical and
numerical analyses, we show that our model can well fit the
current experimental values of three charged lepton masses, two
neutrino mass squared differences and three mixing angles,
although it contains only a few free parameters.

In some recent papers\cite{S4new}, the flavor $S^{}_{4}$ symmetry
has been treated as a subgroup of the continuous flavor $SO(3)$ or
$SU(3)$ symmetry together with the gauge symmetry in some grand
unified theories, i.e. the $SO(10)$ models, and the quark mixing
can be included. A simple way to contain the quark mixing in our
model is to assume that all the quarks belong to the identity {\it
reps}, and then the quark mixing can be obtained via the standard
way. A supersymmetric extension of our model should also be
interesting and may be given elsewhere.

\acknowledgments{The author is indebted to Professor Zhi-zhong
Xing for reading the manuscript, making many corrections and
giving a number of helpful suggestions. The author is also
grateful to Obara Midori, Wei Chao and Shun Zhou for useful
discussions. This work is supported in part by the National
Natural Science Foundation of China.}

\newpage

\appendix{}

\section{representation matrices of $S^{}_4$}

The ${\bf \underline{2}}$ {\it reps} for the class $C^{}_2$ in the
Yamanouchi bases are
\begin{eqnarray}
(12)& \simeq &(34) \sim\left(\matrix{ 1 & 0 \cr 0 & -1 }\right) \
, \ \ \ \ \ (23)\simeq(14) \sim \left(\matrix{ -\frac{1}{2} &
\frac{\sqrt{3}}{2} \cr \frac{\sqrt{3}}{2} & \frac{1}{2} }\right) \
, \nonumber \\
(13)& \simeq &(24) \sim \left(\matrix{ -\frac{1}{2} &
-\frac{\sqrt{3}}{2} \cr -\frac{\sqrt{3}}{2} & \frac{1}{2} }\right)
\ ,
\end{eqnarray}
where the definition of $C^{}_2$ can be found in Table 1. All the
other group elements can be obtained by using the relation
\begin{eqnarray}
(i^{}_1 i^{}_2\cdots i^{}_k \cdots i^{}_{n-1} i^{}_n)=(i^{}_1
i^{}_2 \cdots i^{}_k)(i^{}_k \cdots i^{}_{n-1} i^{}_n) \ .
\end{eqnarray}
The matrices of the ${\bf \underline{3}}_\alpha$ {\it reps} are
given by
\begin{eqnarray}
(12) & \sim & \left(\matrix{  1 & 0 & 0 \cr 0 & 1 & 0 \cr 0 & 0 &
-1} \right) \ , \ \ \ \ \ (23) \sim \left(\matrix{  1 & 0 & 0 \cr
0 & -\frac{1}{2} & \frac{\sqrt{3}}{2} \cr 0 & \frac{\sqrt{3}}{2} &
\frac{1}{2}} \right) \ , \nonumber \\
(34) & \sim & \left(\matrix{ -\frac{1}{3} & \frac{\sqrt{8}}{3} & 0
\cr \frac{\sqrt{8}}{3} & \frac{1}{3} & 0 \cr 0 & 0 & 1} \right) \
, \ \ \ \ \ (13) \sim \left(\matrix{  1 & 0 & 0 \cr 0 &
-\frac{1}{2} & -\frac{\sqrt{3}}{2} \cr 0 & -\frac{\sqrt{3}}{2} &
\frac{1}{2}} \right) \ , \nonumber \\
(14) & \sim & \left(\matrix{ -\frac{1}{3} & -\frac{\sqrt{2}}{3} &
-\sqrt{\frac{2}{3}} \cr -\frac{\sqrt{2}}{3} & \frac{5}{6} &
-\frac{1}{2\sqrt{3}} \cr -\sqrt{\frac{2}{3}} & -\frac{1}{2\sqrt{3}}
& \frac{1}{2} } \right) \ , \ \ \ \ \ (24) \sim \left(\matrix{
-\frac{1}{3} & -\frac{\sqrt{2}}{3} & \sqrt{\frac{2}{3}} \cr
-\frac{\sqrt{2}}{3} & \frac{5}{6} & \frac{1}{2\sqrt{3}} \cr
\sqrt{\frac{2}{3}} & \frac{1}{2\sqrt{3}} & \frac{1}{2} } \right) \ ,
\end{eqnarray}
and the representation matrices in the ${\bf \underline{3}}_\beta$
are
\begin{eqnarray}
(12) & \sim & \left(\matrix{  1 & 0 & 0 \cr 0 & -1 & 0 \cr 0 & 0 &
-1} \right) \ , \ \ \ \ \ (23) \sim \left(\matrix{  -\frac{1}{2} &
\frac{\sqrt{3}}{2} & 0 \cr \frac{\sqrt{3}}{2} & \frac{1}{2} & 0 \cr
0 & 0 & -1} \right) \ , \nonumber \\
(34) & \sim & \left(\matrix{ -1 & 0 & 0 \cr 0 & -\frac{1}{3} &
\frac{\sqrt{8}}{3} \cr 0 & \frac{\sqrt{8}}{3} & \frac{1}{3} }
\right) \ , \ \ \ \ \ (13) \sim \left(\matrix{ -\frac{1}{2} &
-\frac{\sqrt{3}}{2} & 0 \cr -\frac{\sqrt{3}}{2} &
 \frac{1}{2} & 0 \cr 0 & 0 & -1} \right) \ , \nonumber \\
(14) & \sim & \left(\matrix{ -\frac{1}{2} & -\frac{1}{2\sqrt{3}} &
\sqrt{\frac{2}{3}} \cr -\frac{1}{2\sqrt{3}} & -\frac{5}{6}  &
-\frac{\sqrt{2}}{3} \cr \sqrt{\frac{2}{3}} & -\frac{\sqrt{2}}{3} &
\frac{1}{3} } \right) \ , \ \ \ \ \ (24)  \sim  \left(\matrix{
-\frac{1}{2} & \frac{1}{2\sqrt{3}} & -\sqrt{\frac{2}{3}} \cr
\frac{1}{2\sqrt{3}} & -\frac{5}{6}  & -\frac{\sqrt{2}}{3} \cr
-\sqrt{\frac{2}{3}} & -\frac{\sqrt{2}}{3} & \frac{1}{3} } \right) \
,
\end{eqnarray}
The other group elements can be obtained by using Eq. (A2).

\newpage

\begin{table}
\caption{The character table of $S^{}_4$. Here $N$ stands for the
number of elements in class $C^{}_i$, and $C$ denotes the cycle
structure of each class. }
\begin{center}

\begin{tabular}{|c|c|c|c|c|c|c|c|}
 ~~Class~~ & ~~$N$~~ & ~~$C$ ~~& ~~${\bf \underline{1}}_S$~~&~~ ${\bf \underline{1}}_A$~~
 & ~~${\bf \underline{2}}$ ~~& ~~${\bf \underline{3}}_\alpha$ ~~&~~ ${\bf
 \underline{3}}_\beta$~~ \cr
\hline
$C^{}_1$ & 1 & $1^4$ & 1 & 1 & 2 & 3 & 3 \cr
$C^{}_2$ & 6 & $21^2$ & 1 & $-1$ & 0 & 1 & $-1$ \cr
$C^{}_3$ & 8 & $31$ & 1 & 1 & $-1$ & 0 & 0 \cr
$C^{}_4$ & 6 & $4$ & 1 & $-1$ & 0 & $-1$ & 1 \cr
$C^{}_5$ & 3 & $2^2$ & 1 & 1 & 2 & $-1$ & $-1$ \cr
\end{tabular}
\vspace{2cm}
\caption{The Higgs potential at its minimum. The parameters $K_i
(i=1-4)$ are defined as $K_1=\kappa_1 + 4\kappa_3 + 4\kappa_4 +
2\kappa_5 + 2\kappa_6$, \ \ \ $K_2=\kappa_1 - 2\kappa_2 + 2\kappa_3
+ 6\kappa_4 + \kappa_5 + 3\kappa_6$, \ \ \ $K_3=\kappa_1 + 2\kappa_2
+ 6\kappa_3 + 2\kappa_4 + 3\kappa_5 + \kappa_6$ \ \ \ and \ \ \
$K_4= 2\kappa_2 + 2\kappa_3 -2\kappa_4 +\kappa_5 - \kappa_6$.}
\vspace{0.5cm}
\begin{tabular}{|c|c||c|c|}
~~~~~~~~~~~~~~Terms~~~~~~~~~~ & Coefficients & Terms & Coefficients
\cr \hline
$v^2_u$ & $-m^2_u$ & $v^2_d$ & $-m^2_d$ \cr
\hline $ u^2_1 + u^2_2$ & $-m^2_\phi$ & $w^2_1+w^2_2+w^2_3$ &
$-m^2_\chi$ \cr
\hline $ v^4_u$ & $\lambda_u$ & $v^4_d$ & $\lambda_d$ \cr
\hline $ (u^2_1 + u^2_2)^2 $ & $\lambda_\phi + \tau_1$ & $(w^2_1 +
w^2_2 + w^2_3)^2 $ & $\lambda_\chi$ \cr
\hline  $w^2_1$ & $4\mu_1$ & $w^4_2+w^4_3$ & $3\mu_1+\mu_3$ \cr
\hline $w^2_1 w^2_2 + w^2_1 w^2_3$ & $ 8 \mu_3$ & $w^2_2 w^2_3$ & $
6 \mu_1+2\mu_3$ \cr
\hline $ w_1 w^3_2 $ & $4\sqrt{2}(\mu_3 - \mu_1)$ & $w_1 w_2 w^2_3$
& $12\sqrt{2}(\mu_1 - \mu_3)$ \cr
\hline $v^2_u (u^2_1+u^2_2)$ &
$\lambda^\phi_1+2\lambda^\phi_2+\lambda^\phi_3$ & $v^2_d
(u^2_1+u^2_2)$ & $\sigma^\phi_1+2\sigma^\phi_2+\sigma^\phi_3$ \cr
\hline $v^2_u (w^2_1+w^2_2+w^2_3)$ &
$\lambda^\chi_1+2\lambda^\chi_2+\lambda^\chi_3$ & $v^2_d
(w^2_1+w^2_2+w^2_3)$ &
$\sigma^\chi_1+2\sigma^\chi_2+\sigma^\chi_3$\cr
\hline $v^2_u v^2_d$ & $\lambda_1+\lambda_2+2\lambda_3$ & $v_u
u^3_1$ & $2\lambda^\phi_4$ \cr
\hline $v_u u_1 u^2_2$ & $-6\lambda^\phi_4$ & $v_d w^3_1$ &
$4\sigma^\chi_4$ \cr
\hline $v_d w^3_2$ & $2\sqrt{2}\sigma^\chi_4$ & $v_d w_1 (w^2_2 +
w^2_3) $ & $-6 \sigma^\chi_4$ \cr
\hline $v_d w_2  w^2_3$ & $-6 \sqrt{2} \sigma^\chi_4$ &
$(u^2_1+u^2_2) w^2_1$ & ${K}_1$ \cr
\hline $u^2_1 w^2_3 + u^2_2 w^2_2 $ & $K_2$ & $u^2_1 w^2_2 + u^2_2
w^2_3 $ & $K_3$  \cr
\hline $u^2_1 w_1 w_2$ & $ 2\sqrt{2}K_4$ & $u_1 u_2 w_1 w_3$ & $
-4\sqrt{2}K_4$  \cr
\hline $u_1 u_2 w_2 w_3$ & $ 4 K_4$ & $u^2_2 w_1 w_2$ & $-2\sqrt{2}
K_4$ \cr
\hline $v_u u_1 w_1 w_2$ & $4\sqrt{2} (\delta_1+\delta_2+\delta_3)$
& $v_u u_1 w^2_2$ & $2(\delta_1 + \delta_2+\delta_3)$ \cr
\hline $v_u u_1 w^2_3$ & $-2(\delta_1+\delta_2+\delta_3)$ & $v_u u_2
w_1 w_3$ & $4\sqrt{2} (\delta_1+\delta_2+\delta_3)$ \cr
\hline $v_u u_2 w_2 w_3$ & $-4 (\delta_1+\delta_2+\delta_3)$ & &
\end{tabular}

\end{center}
\end{table}
\begin{figure}
\psfig{file=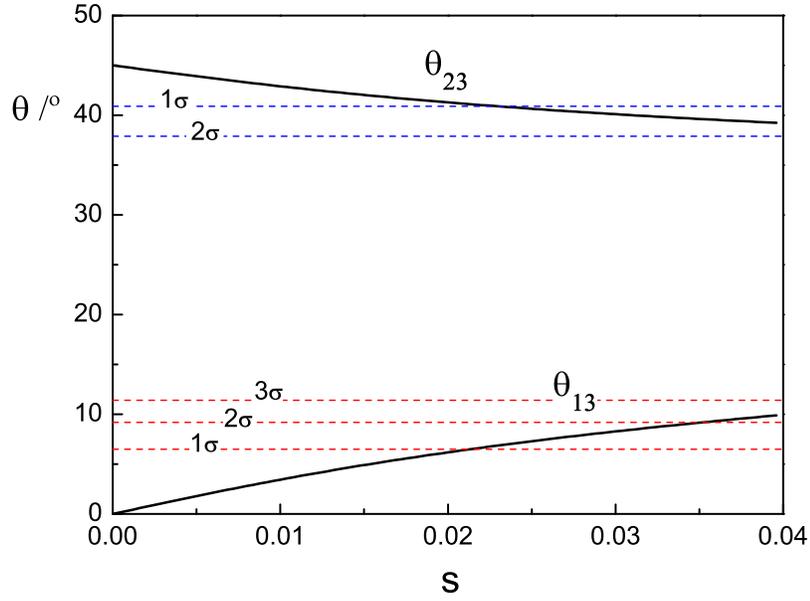, bbllx=-1cm, bblly=-1.0cm, bburx=9cm, bbury=9.0cm,%
width=11cm, height=11cm, angle=0, clip=0}\vspace{-1.3cm}
\caption{Relations between nonzero $s$ and mixing angles
$(\theta_{13},\theta_{23})$. The dashed lines correspond to the 1, 2
and 3$\sigma$ confidence level from oscillation experiments. We also
take $\Delta m^2_{32}=2.5\times 10^{-3} {\rm eV}^2$, $\Delta
m^2_{21}=8.0\times 10^{-5} {\rm eV}^2$ and
$\theta_{12}=33.9^{\circ}$ in our calculations.}
\end{figure}

\end{document}